\documentclass[lettersize,journal]{IEEEtran}\usepackage{graphicx}
\usepackage{amsmath,amsfonts}
\usepackage{algorithmic}
\usepackage{algorithm}
\usepackage{array}
\usepackage[caption=false,font=normalsize,labelfont=sf,textfont=sf]{subfig}
\usepackage{textcomp}
\usepackage{stfloats}
\usepackage{url}
\usepackage{verbatim}
\usepackage{graphicx}
\usepackage{cite}
\usepackage{color}
\usepackage{enumitem}
\usepackage{pifont}
\newcounter{circnum}
\usepackage[table,xcdraw]{xcolor}
\usepackage{fontawesome5}

\begin{document}

\title{NetOrchLLM: Mastering Wireless Network Orchestration with Large Language Models}

\author{
Asmaa Abdallah$^\star$, Abdullatif Albaseer$^\star$, Abdulkadir Celik, Mohamed Abdallah, Ahmed M. Eltawil
\vspace{-0.75cm}
\thanks{$^\star$ The first two authors contributed equally to this work.}
}




\maketitle

\begin{abstract}
The transition to 6G networks promises unprecedented advancements in wireless communication, with increased data rates, ultra-low latency, and enhanced capacity. However, the complexity of managing and optimizing these next-generation networks presents significant challenges. 
The advent of large language models (LLMs) has revolutionized various domains by leveraging their sophisticated natural language understanding capabilities. However, the practical application of LLMs in wireless network orchestration and management remains largely unexplored. Existing literature predominantly offers visionary perspectives without concrete implementations, leaving a significant gap in the field. 
To address this gap, this paper presents \textsc{NetOrchLLM}, a wireless \underline{\textsc{Net}}work \underline{\textsc{Orch}}estrator \underline{\textsc{LLM}} framework that uses LLMs to seamlessly orchestrate diverse wireless-specific models from wireless communication communities using their language understanding and generation capabilities. 
A comprehensive framework 
is introduced, demonstrating the practical viability of our approach and showcasing how LLMs can be effectively harnessed to optimize dense network operations, manage dynamic environments, and improve overall network performance. \textsc{NetOrchLLM} bridges the theoretical aspirations of prior research with practical, actionable solutions, paving the way for future advancements in integrating generative AI technologies within the wireless communications sector. 
\end{abstract}

\section*{\textbf{Introduction}}
The symbiosis of 6G and artificial intelligence (AI) promises a transformative era by making machine learning (ML) routines native to  wireless network hardware, design, optimization, and operation \cite{rong2024leveraging}. This paradigm shift towards AI-native generations is poised to revolutionize various facets of wireless networks, from resource allocation and interference management to network management and orchestration. 
Following decades of continuous efforts highlighting the advantages of discriminative AI over model-based approaches, the wireless networking research community has recently turned its attention to generative AI (GenAI), driven by its remarkable success in various domains such as vision, natural language processing, speech synthesis, etc.  Various generative models have shown promising results for wireless network problems, such as physical layer design, resource allocation, network traffic analytics, cross-layer security, and localization~\cite{celik2024genai}. 

Among the array of GenAI technologies, large language models (LLMs) are particularly noteworthy. LLMs have recently been applied in a spectrum of telecom applications to enable intelligent interaction within network management systems \cite{zou2023wireless,xu2024large,shen2024large,bariah2024large,jiang2024large,shao2024wirelessllm,tarkoma2023ai,6Gllm,yilma2024telecomrag,maatouk2023teleqna,zou2024telecomgpt,nazar2024enwar}. Their ability to process and generate context-aware, text-based responses makes them promising for enhancing operational efficiency and responsiveness across the network and physical layers. 
Integrating LLMs into telecommunication infrastructure still faces several limitations, including susceptibility to hallucinations, being treated as chatbots, and reliance on static, outdated knowledge bases.  
While LLMs excel in natural language processing, they often struggle with the technical complexities of wireless communication, such as dynamic resource allocation, network management, and traffic pattern analysis, tasks that demand advanced mathematical computations and require specialized models to handle sub-tasks across multiple systems. Although LLMs show promising zero-shot or few-shot learning capabilities, they still fall short of expert models in handling more complex, real-world challenges. Furthermore, next-generation wireless networks are expected to leverage multi-modal communication (RSSI, I/Q data, channel matrices, etc.) and sensory (camera, LiDAR, radar, GPS, etc.) data modalities.  These diverse data types have often been overlooked in existing LLM-focused frameworks, limiting their effectiveness in real-world multi-modal settings.

In this article, {we address key limitations in existing LLM-driven wireless approaches by introducing  \textsc{NetOrchLLM}, a wireless \underline{\textsc{Net}}work \underline{\textsc{Orch}}estrator \underline{\textsc{LLM}} framework developed for large-scale networks and equipped with an inter-model cooperation protocol.} \textsc{NetOrchLLM} leverages LLMs as the central intelligence for
strategic planning and decision-making by automatically deploying expert models tailored to specific
tasks. The proposed framework bridges theoretical research with practical implementation, paving the way for efficient, scalable, and adaptable network management solutions. The paper provides a comprehensive overview of LLM-supported wireless networks to identify critical limitations in current methodologies and to highlight the relevance of \textsc{NetOrchLLM}’s contributions. Next, we outline \text{NetOrchLLM} framework and present two exemplary case studies—focusing on bandwidth and power allocation—that demonstrate \textsc{NetOrchLLM}'s effectiveness in addressing complex problems and surpassing the performance of traditional LLMs, especially as user and cell scales increase. 

\section*{\textbf{State-of-the-Art LLM Perspectives, \\ Common Limitations, and Open Challenges}}

This section explores current perspectives on how LLM capabilities are leveraged to support wireless communication and highlights the limitations of existing research methodologies.

\subsection*{\textbf{State-of-the-Art LLM Perspectives}}

\begin{itemize}
    \item \textbf{Retrieval-Augmented Generation (RAG)} is a framework designed to extend LLM capabilities by incorporating domain-specific external knowledge into the response generation process. In this architecture, a domain-specific dataset is first curated and embedded into a vector space to create a knowledge base. Likewise, input queries are also vectorized through embedding and used to perform a semantic search across the knowledge base. The retrieved information is then ranked based on relevance, and the most pertinent snippets are provided as context for the model response generation \cite{xu2024large,yilma2024telecomrag,nazar2024enwar}. This process enables RAG models to deliver more precise and contextually appropriate answers, especially for complex queries. In the context of wireless communication, RAG can significantly improve the model’s handling of technical queries by accessing historical data, pre-trained models tailored for specific tasks and protocols, or industry standards directly from the knowledge base, thereby ensuring accuracy and relevance in responses.

    \item \textbf{Question and Answer (Q\&A) Training} creates multiple choice question (MCQ) benchmarks from a large corpus of documents in the telecom domain \cite{maatouk2023teleqna}. It encompasses tasks such as question answering, where all correct answers are selected from MCQs, and telecom-related questions are answered based on standards, research papers, or patents.
    
    \item \textbf{Telecom-Specific Instruct and Alignment Tuning} are two effective ways of improving LLM performance \cite{zou2024telecomgpt,tarkoma2023ai}. To improve instruction-following and task handling, supervised fine-tuning is applied, enhancing their ability to perform well in zero-shot or few-shot scenarios and reducing refusal rates. 
 Additionally, alignment tuning refines LLM responses to match human preferences. While reinforcement learning with human feedback is commonly used, it can be costly and unstable due to the complexity of preference data and training objectives. An alternative approach, direct preference optimization using a Bradley-Terry model, simplifies the process by avoiding explicit reward functions and improving response accuracy.
    
    \item \textbf{Multi-Agent Collaboration} can be leveraged to optimize task-solving capabilities in 6G networks \cite{shen2024large,zou2023wireless,bariah2024large,jiang2024large}, wherein users can express their task requirements in natural language. Multi-agent data retrieval may query and summarize domain-specific knowledge in 6G communications from private data. The collaborative planning can decompose the original task based on the retrieved communication knowledge, generate multiple feasible sub-task chains, and solve them. 
\end{itemize}

\subsection*{\textbf{Common Limitations and Open Challenges}}
Despite the valuable contributions to date, these early efforts in utilizing LLMs for wireless networks share the following notable limitations:
\subsubsection*{\textbf{Treating LLMs as Query Tools}}
While Q\&A techniques treat LLMs as advanced query tools or "GPT bots," performing tasks such as knowledge queries, math modeling, Tdocs classification, and code generation and analysis in a telecom context, they fall short of solving intricate wireless communication problems. These methods primarily leverage LLMs for extracting and summarizing information, rather than addressing the complex and dynamic nature of wireless networks. Our approach, in contrast, aims to harness LLMs for advanced problem-solving in wireless communication by coordinating with external models, designing RF precoders, optimizing resource allocation, and managing multi-user interference. This involves not only information retrieval but also the integration and application of multi-modal data to enhance the efficiency and efficacy of wireless communication networks.

\subsubsection*{\textbf{Inability to Address NP-Hard and Large-Scale Problems}}
While current approaches \cite{zou2023wireless,xu2024large,shen2024large,bariah2024large,jiang2024large,shao2024wirelessllm,tarkoma2023ai,6Gllm,yilma2024telecomrag,maatouk2023teleqna,zou2024telecomgpt,nazar2024enwar} demonstrate how LLMs can be leveraged to solve basic wireless communication tasks, they fall short when faced with more challenging, non-convex, or NP-hard problems. 
As the complexity of the problem increases, particularly in scenarios involving large-scale networks or multi-user environments, LLMs struggle to generate accurate or optimal solutions. 
As LLMs are adept at handling simpler tasks or small-scale problems, they lack the specialized optimization techniques and mathematical rigor needed for solving highly complex and large-scale wireless communication challenges. More advanced methods, such as hybrid models combining LLMs with optimization solvers or domain-specific algorithms, is required to address these limitations and extend the capabilities of LLMs in handling real-world, large-scale, non-convex optimization problems.

\subsubsection*{\textbf{Lack of Multi-Modality}} Existing research mainly relies on text data, which is insufficient for the complex, multi-modal nature of wireless communication tasks that involve sensory data, physical signals, radar, LiDAR, and camera inputs. Tasks like beamforming require specific signal data beyond what text can convey. Effective wireless models need to integrate diverse data sources such as signal strength, interference, and environmental context for real-time optimization. Although the multi-modal potential is discussed in \cite{xu2024large,shao2024wirelessllm,bariah2024large,6Gllm}, these works lack concrete case studies or proof-of-concept implementation. While \cite{nazar2024enwar} includes a case study, it only describes environments visually and lacks problem-solving capabilities essential in wireless communications.

\subsubsection*{\textbf{Static and Outdated Knowledge}}
Knowledge bases used in RAG models are often static, requiring regular updates to remain effective. In the fast-evolving field of wireless communication, outdated information can lead to incorrect or suboptimal decisions, particularly in dynamic scenarios or when new standards, technologies, or practices emerge. While RAG provides promising results by integrating external knowledge for improved contextual responses, it still depends on the timeliness and relevance of the knowledge base. To maintain accuracy and effectiveness, RAG-enabled LLMs must be connected to continuously updated knowledge sources, ensuring the model can adapt to changes and provide reliable insights in real-time.

\subsubsection*{\textbf{Generation of Plausible yet Incorrect Information}}

LLMs may occasionally produce responses that sound plausible but are factually incorrect, a phenomenon known as "hallucination." In the field of wireless communication, this issue can have serious implications, as inaccurate information might lead to critical misconfigurations, network failures, or security vulnerabilities. Thus, it is essential to ensure the reliability and accuracy of LLM outputs, particularly in high-stakes environments where precision is crucial. 

\begin{figure*}[t]
     \centering
\includegraphics[width=.85\linewidth]{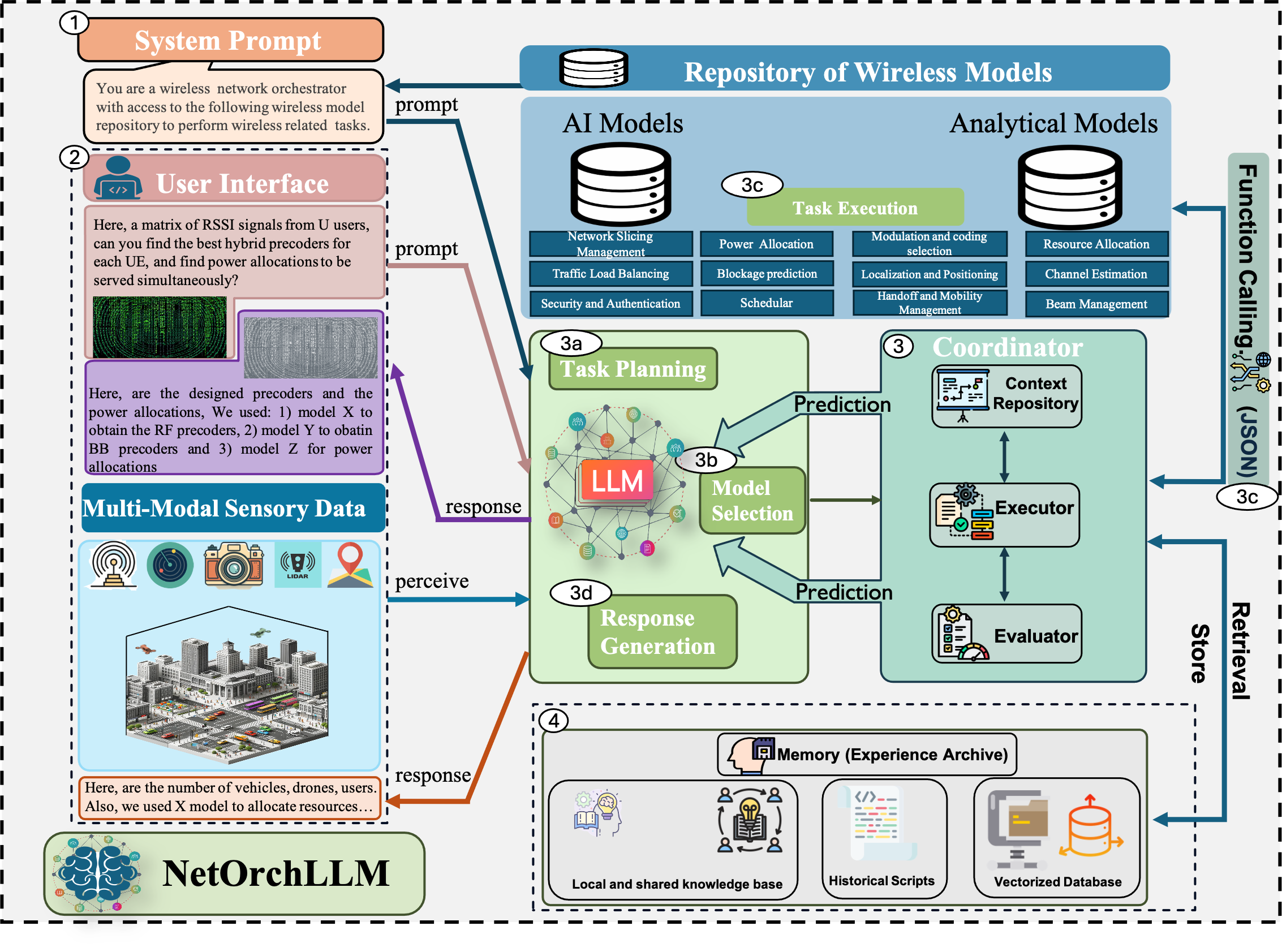}
     \caption{{The overall framework of \textsc{NetOrchLLM}.}}
    \label{fig:system_arch}
\end{figure*}

\section*{\textbf{An Overview of NetOrchLLM}}

As LLMs face limitations in addressing the complex, multi-task demands of wireless communication, such as network and resource management due to their single-model design, real-world wireless communication tasks are often multifaceted, requiring the coordination of multiple sub-tasks and the collaboration of various models. This necessity for multi-model cooperation extends beyond the standalone capabilities of current LLMs. 
To overcome these challenges, we propose utilizing LLMs as a central orchestrator, referred to as \textsc{NetOrchLLM}, which manages planning, scheduling, and collaboration with specialized AI models through a language-based interface. This approach allows LLMs to delegate intricate tasks, such as network management and resource allocation, to domain-specific models, enhancing the LLM’s ability to handle complex wireless communication problems. By integrating well-defined analytical and/or AI models, \textsc{NetOrchLLM} enhances LLM capabilities to provide scalable solutions for wireless network orchestration. The following sections detail the architecture and key components of the \textsc{NetOrchLLM} framework, as illustrated in Fig. \ref{fig:system_arch}.

\subsection*{\hspace{10pt}\textbf{Fitting LLMs for Wireless Communication Tasks}}

\textsc{NetOrchLLM} expands LLMs' wireless communication capabilities by incorporating a repository of specialized models, knowledge-augmented generation techniques, and multi-modal training methods, addressing inherent limitations in traditional LLM applications.

\subsubsection*{\hspace{-15pt} \textcircled{\textcircled{$\star$}} 
\textbf{Analytical and Data-Driven Wireless Model Repository}} 
In order to excel in performing wireless communication tasks, \textsc{NetOrchLLM} utilizes a model repository, wherein each model is paired with relevant tasks. This repository, as illustrated in Fig. \ref{fig:system_arch}, provides a structured approach to handle diverse tasks such as resource allocation, channel estimation, and beam prediction. Each task is mapped to a model optimized for that function, ensuring that \textsc{NetOrchLLM} can efficiently manage complex communication scenarios. The challenge lies in addressing a wide range of AI tasks, which demands detailed and high-quality model descriptions. This is achieved through prompt engineering, a key component in accurately representing user requests and mapping them to the appropriate models. By incorporating structured model descriptions into prompts, \textsc{NetOrchLLM} ensures effective collaboration between the LLM and domain-specific analytical and AI models. For instance, prompts can request AI-Channel Estimation or AI-beam prediction models, and the system, using these descriptions, identifies and invokes suitable models. The process is further streamlined by employing slot-filling methodologies. Slot-filling ensures that tasks are represented precisely, filling in key parameters such as RSSI signals or UE requirements, as depicted in the user interface and task planning modules in Fig. \ref{fig:system_arch}. This structured approach facilitates seamless model integration, allowing \textsc{NetOrchLLM} to orchestrate tasks efficiently and generate accurate responses.

\subsubsection*{\hspace{-15pt} \textcircled{$\star$} 
 \textbf{Knowledge-Augmented Generation}}
To address challenges such as hallucination and outdated knowledge, our framework supports RAG integration to strengthen the \textsc{NetOrchLLM}’s knowledge base. Similar to \cite{xu2024large,yilma2024telecomrag,nazar2024enwar}, RAG can integrate a variety of external, authoritative resources—such as research publications, device manuals,  textbooks, and industry standards to ensure \textsc{NetOrchLLM} has access to up-to-date, reliable information. This allows \textsc{NetOrchLLM} to produce contextually relevant, accurate responses for complex, knowledge-intensive tasks such as link adaptation and modulation schemes. Additionally, \textsc{NetOrchLLM}  continuously refines its outputs by incorporating new data and insights from user interactions.

As outputs are generated through the repository of analytical and data-driven models, they are stored within a dedicated and evolving knowledge base, forming a proprietary reservoir of insights. This curated knowledge base allows \textsc{NetOrchLLM} to reference past responses for similar prompts and scenarios over time, potentially bypassing the need to re-run certain models, are stored in a memory as highlighted in Fig. \ref{fig:system_arch} and explained in the sequel.  

\subsubsection*{\hspace{-15pt} \textcircled{\textcircled{$\star$}} 
 \textbf{Multi-Modal Sensory Data}}

In our \textsc{NetOrchLLM} framework, we build on our previous work with ENWAR \cite{nazar2024enwar}, an ENvironment-aWARe, RAG-empowered multi-modal LLM specifically designed to handle multi-modal sensory data for enhanced environmental perception. ENWAR integrates various data modalities such as GPS, LiDAR, and camera data to create a rich, human-interpretable situational awareness in complex wireless environments. By preprocessing and transforming these sensory inputs into a unified text-based format, ENWAR allows \textsc{NetOrchLLM} to perceive, interpret, and generate accurate responses that reflect the real-world spatial dynamics of the environment. Incorporating domain-specific datasets is essential for maximizing the effectiveness of LLMs in wireless communication. Multi-modal pre-training enhances model performance by leveraging various data types relevant to the wireless domain, including text, physical symbols, signal patterns, and environmental noise. Independent encoders process and extract features from each modality, which are then tokenized and passed through a transformer architecture. This approach allows \textsc{NetOrchLLM} to understand and establish correlations across different data types, thereby enhancing its ability to handle complex wireless communication tasks. With access to a repository of AI models tailored for multi-modal data processing, \textsc{NetOrchLLM} effectively supports multimodal fusion, integrating diverse sensory inputs to deliver more accurate and contextually aware responses.

\subsection*{\hspace{10pt} \textbf{\textsc{NetOrchLLM} Framework Breakdown}}
The framework integrates LLMs into wireless network orchestration, leveraging their advanced natural language processing capabilities to enhance network management. The following sections detail the architectural elements within the framework, illustrated in Fig. \ref{fig:system_arch}. 

\subsubsection*{{\textcircled{\footnotesize \textbf{1}}} \textbf{System Prompt Initialization}}
The process begins with the \textit{System Prompt}, which is used to instruct \textsc{NetOrchLLM} about its role in the system. In this prompt, the LLM is informed that it has access to a repository of wireless-specific analytical and AI models that can be called upon as needed. This is the first layer where the system understands its environment and the resources it can use.

\subsubsection*{\textcircled{\footnotesize \textbf{2}} \textbf{User Query Input and Multimodal Data}}
The user can interact with \textsc{NetOrchLLM} through the {user interface} to which the user submits queries including data related to wireless signals such as RSSI, power levels, or beamforming matrices or environmental data from external sensors, such as GPS, LiDAR, or camera data. This multimodal input is crucial for the system to process real-time data from complex environments, such as smart cities or autonomous networks.
\subsubsection*{\textcircled{\footnotesize \textbf{3}} \textbf{Coordinator and Task Execution}}
The \textit{Coordinator} is essential for orchestrating the tasks within the framework, ensuring below steps are efficiently executed.

\begin{enumerate}[label=\textcircled{\footnotesize \textbf{\alph*}}]
    \item \textbf{Task planning:} 
\textsc{NetOrchLLM}  analyzes the query and breaks it down into smaller, structured tasks. For instance, if the query involves optimizing power allocation and channel estimation, the system splits these into independent tasks. The LLM’s capability to understand natural language allows it to decompose the user request into manageable sub-tasks for further execution.

\item \textbf{Model selection:}
Based on the user’s query and the specific tasks identified, \textsc{NetOrchLLM} selects the most suitable models from the repository to address each task efficiently. Model selection is conducted through a dynamic in-context task-model assignment mechanism, where the available models are matched to tasks by utilizing detailed model descriptions as a language interface.  
\textsc{NetOrchLLM} first filters models by task type, ensuring that only relevant models are considered for the current task. The system then ranks these filtered models based on predefined criteria—such as, cosine similarity, download frequency or relevance—to select the top candidates for execution. By narrowing down and prioritizing models in this way, \textsc{NetOrchLLM} reduces token usage in prompts and maximizes the efficiency of task execution.
\item \textbf{Function calling and task execution:}
\textsc{NetOrchLLM} uses JSON-formatted functions to interact with these models, allowing it to execute tasks such as determining the optimal power allocation, predicting beam directions, or estimating channels. During this phase, \textsc{NetOrchLLM} works closely with the selected models to solve the user's request by utilizing advanced algorithms for optimal performance.
\item \textbf{Response generation}
\textsc{NetOrchLLM} synthesizes the outputs from the selected models and generates a comprehensive response for the user. The LLM’s NLP capability enables it to present technical results in a human-interpretable format. {Once the response is generated, the coordinator evaluates the performance of the task and utilizes the feedback mechanism for continuous system improvement. This iterative feedback process is crucial for grounding the system’s decisions in real-world performance metrics, thereby enhancing reliability and user trust in the generated solutions.}
\end{enumerate}
The coordinator also contains a \textit{context repository} for storing and managing contextual information relevant to current network conditions and user queries. It includes both \textit{contextual data}, capturing real-time network status, and \textit{historical context}, storing past network states and user interactions. These elements help in informed decision-making by allowing the system to analyze past trends and anticipate future conditions.

\subsubsection*{\textcircled{\footnotesize \textbf{4}} \textbf{Memory Management}}
\textsc{NetOrchLLM} incorporates a robust {memory management} component, which stores past interactions, results, and performance metrics in a {Memory, i.e., experience archive}. This archive consists of a {local and shared knowledge base} and a {vectorized database} for efficient retrieval of previously learned information. \textsc{NetOrchLLM} uses this memory to retrieve relevant historical data and leverage it for improving future task execution and decision-making. This memory-based system allows for continuous improvement over time, making the system smarter with each interaction. To ensure that  \textsc{NetOrchLLM}  continuous adaptation, user feedback is collected and used to refine future responses and adapt to changing scenarios. This feedback can be used to adjust the models' decision-making process and improve accuracy for similar queries in the future.

\begin{figure*}[!t]
 \centering\includegraphics[width=0.85\linewidth]{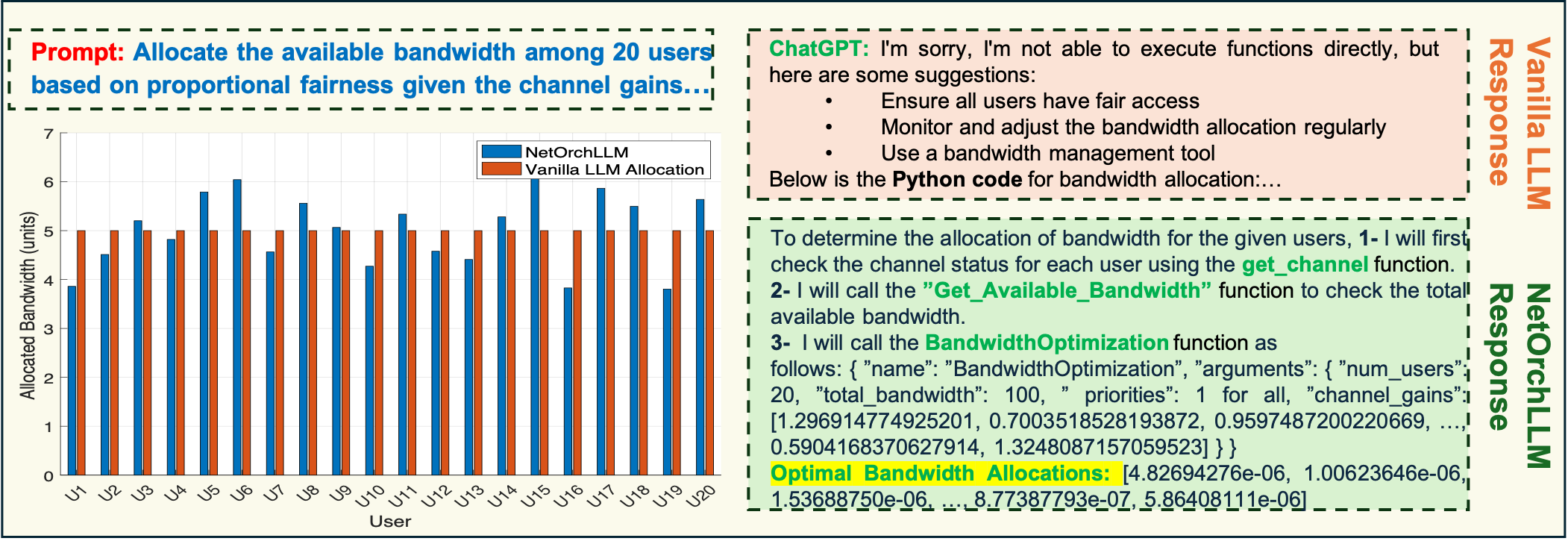}
  \caption{Comparison between baseline LLM responses and \textsc{NetOrchLLM} framework in terms of bandwidth allocation task }
    \label{fig:BA_Scenario}
\end{figure*}

\section*{\textbf{Case Studies and Discussions}}
A performance evaluation is carried out for two case studies on bandwidth and power allocation to demonstrate how \textsc{NetOrchLLM} performs in real-world wireless network environments, particularly focusing on its ability to handle complex tasks and its scalability across diverse network conditions. In both cases, we compared our approach to baseline methods such as ChatGPT 4.o with and without RAG. Notably, \textsc{NetOrchLLM} leverages the lightweight Mistral LLM, which is significantly more resource-efficient than larger models.

\subsubsection*{\textbf{Bandwidth Allocation Scenario}}
The bandwidth allocation case study is conducted for a 6G network environment. The objective is to optimize the distribution of 100 bandwidth units among 20 user equipment (UE) for proportional fairness, given the channel gains of the UEs.

The baseline approach, using ChatGPT 4.o without integrating domain-specific models, struggled with task accuracy. As illustrated in Fig.~\ref{fig:BA_Scenario}, when prompted for bandwidth allocation, the LLM provided general optimization guidelines but failed to offer a concrete solution. Without specialized models, the vanilla LLM generated a Python script that allocated bandwidth equally among users, resulting in inefficient resource utilization. 
In contrast, \textsc{NetOrchLLM} effectively selected the appropriate model from the repository, optimizing bandwidth usage while ensuring fair distribution among users. As shown in Fig.~\ref{fig:BA_Scenario}, the system leveraged real-time environmental data—such as channel gains and interference patterns—to dynamically adjust bandwidth allocation, enabling better adaptation to changing network conditions.
\begin{figure*}
\centering
\includegraphics[width=0.85\linewidth]{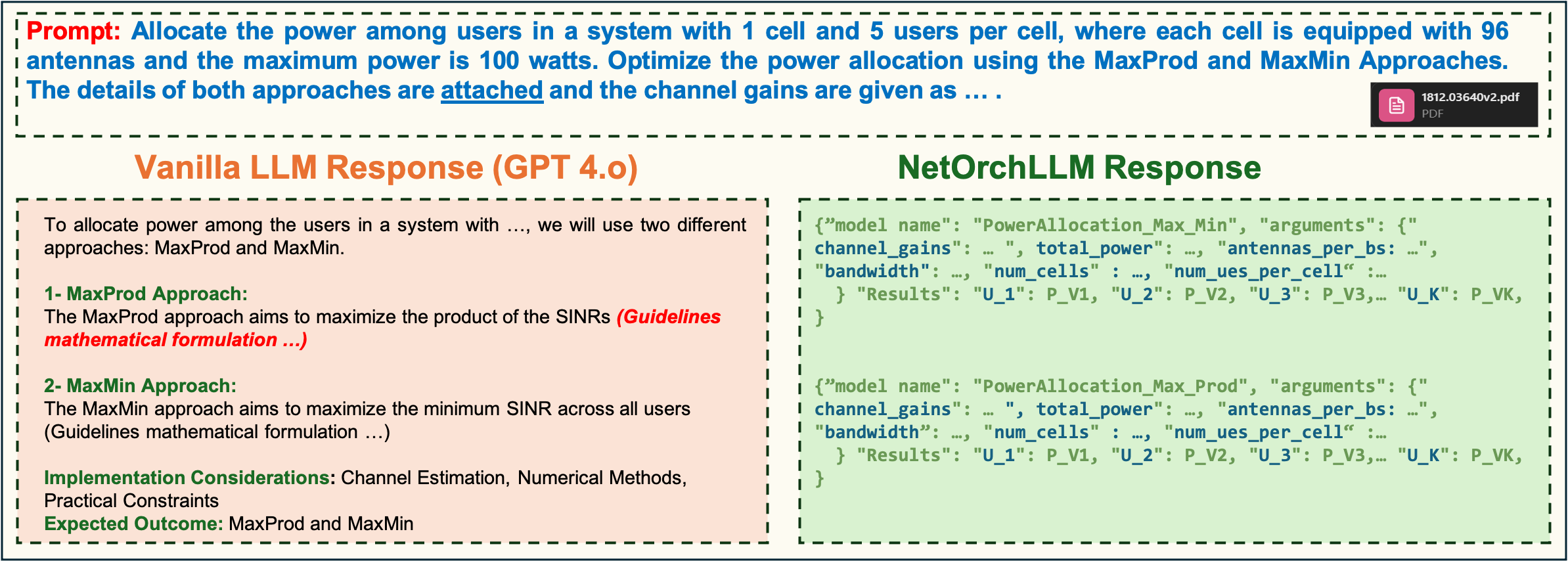}
 \caption{Comparison of baseline and our proposed framework responses to a complex power allocation task.}   
 \label{fig:PA_base_prompt}
\end{figure*}

\subsubsection*{\textbf{Power Allocation Scenario}}
The second case study examined the more complex task of power allocation in a multi-cell massive MIMO system. In this scenario, multiple base stations, each equipped with 96 antenna systems, are connected to $K$ users, making power allocation significantly more challenging than the previous bandwidth allocation scenario.\\
As in Fig. ~\ref{fig:PA_base_prompt}, we initially prompted ChatGPT to perform power allocation, and it provided high-level guidelines, suggesting general optimization techniques even when guided by RAG, and provided manuscript \cite{PA2018debbah} detailing optimization techniques for power allocation in multi-cell massive MIMO systems. We note that  ChatGPT 4.o still offered general guidelines without producing a working solution. This highlights the limitations of the baseline LLM in performing complex optimization tasks that require mathematical precision.
Further attempts involved prompting the LLM to generate an optimization solution. While the LLM produced code to solve the problem as an optimization, it was non-functional and failed to compile. Then, we explicitly prompted the LLM to write the water-filling algorithm for power allocation. The code generated by the LLM compiled and merely distributed power uniformly without accounting for MIMO antennas and inter and intra-cell interference. 
This demonstrated the fundamental limitations of using LLMs for complex, multi-dimensional optimization problems in network management.

\begin{figure*}[!t]
    \centering
\includegraphics[width=1\linewidth]{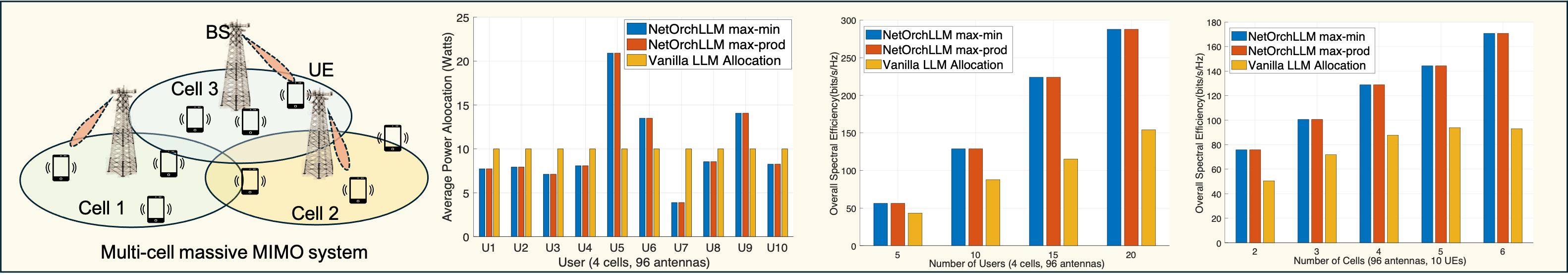}
    \caption{Power allocation using our proposed approach versus vanilla LLM (GPT 4.o).}
    \label{fig:PA}
\end{figure*}

As shown in Fig. \ref{fig:PA_base_prompt}, \textsc{NetOrchLLM} effectively addresses the power allocation problem where the LLM selected the suitable model for the respective optimization objective. In the max-min approach it focused on maximizing the minimum signal-to-interference plus noise (SINR) among all UEs (max-min) to ensure fairness by focusing on the UEs with the worst channel condition. While in the max product approach, it maximized the product of SINRs of the UEs to ensure proportional fairness. \\ 
Our approach exhibits several key strengths. First, it manages interference by accounting for inter-cell and intra-cell interference, ensuring optimal power distribution across all users and antennas in a MIMO-enabled environment. Second, as illustrated in Fig.~\ref{fig:PA}, our system remains scalable regardless of the number of cells or users in the network. As the network size increased, the proposed framework consistently delivered solid results, optimizing power allocation even in large-scale networks. 

\section*{\textbf{Open Problems and Future Directions}}
We have proposed  \textsc{NetOrchLLM} framework as a robust solution for wireless network orchestration, addressing key limitations found in the current landscape of LLM-based applications in telecommunication. \textsc{NetOrchLLM} integrates LLMs with a repository of specialized models to manage complex network tasks, such as resource allocation, power management, and bandwidth optimization, in dynamic and large-scale network environments. Through case studies on bandwidth and power allocation, we demonstrated the advantages of our approach over standard LLMs, highlighting improvements in efficiency, scalability, and adaptability.

While our LLM-based framework for wireless network orchestration has demonstrated significant potential, several technical challenges and future research directions need to be addressed, including:
\subsubsection*{\textbf{Baby LLMs and Federated Learning}}
Deploying LLM agents across various network layers, including cloud-based LLMs and edge-based "\textit{baby LLMs}," can improve efficiency and adaptability in wireless communication systems. A distributed and federated LLM architecture, which integrates smaller, lightweight baby LLMs at the edge with full-scale cloud LLMs, optimizes both real-time performance and scalability. Baby LLMs handle critical tasks using techniques like model pruning and quantization, reducing dependence on cloud resources, while more complex computations are offloaded to cloud-based LLMs. Further performance improvements can be achieved by enabling baby LLMs to process sensory data directly at the edge, forming local RAGs and reducing latency. This architecture requires cooperative knowledge base optimization and memory management to ensure seamless interaction between LLMs, allowing them to collaborate and share insights. However, challenges arise from the differing capabilities of resource-constrained edge devices and cloud environments, underscoring the need for adaptable, non-uniform solutions to balance efficiency and performance across network layers.

\subsubsection*{\textbf{Data Scarcity and Database Limitations}}
A major challenge is the scarcity of extensive and diverse datasets necessary to train robust LLMs for wireless communication. The available data might not encompass all scenarios or variations encountered in real-world applications, leading to gaps in the model's knowledge. Additionally, wireless communication tasks often require highly specialized databases containing signal patterns, network configurations, and performance metrics. These databases are scarce and challenging to compile, limiting the breadth of training data.
 
\subsubsection*{\textbf{Real-Time Adaptability and Responsiveness}}
Real-time adaptability is crucial for managing dynamic and unpredictable wireless network environments. 
Enhancing the framework's ability to perform \textit{low-latency inference} is also essential to ensure timely responses to critical network events and user requests. Thus, optimizing model architectures and leveraging hardware accelerators such as GPUs, TPUs, and specialized AI chips can significantly reduce inference times. 
\subsubsection*{\textbf{Security and Privacy Considerations}}
As LLMs are integrated into wireless network orchestration, ensuring the security and privacy of network operations becomes paramount. Future research should address potential vulnerabilities and develop robust security measures.
Establishing \textit{secure model deployment} methods is essential to protect LLMs from potential attacks and unauthorized access. Implementing encryption protocols, secure boot mechanisms and hardware security modules can enhance the security of model deployment. Secure multi-party computation techniques can be utilized to perform computations on encrypted data without revealing the data itself.

\subsubsection*{\textbf{Advanced Algorithmic Developments}}
Integrating \textit{reinforcement learning} approaches enables LLMs to learn optimal policies for network management tasks through interaction with the environment. Techniques such as deep reinforcement learning can improve the system's ability to make decisions in uncertain and dynamic contexts, adapting to changes in network conditions and user demands.

\subsubsection*{\textbf{Energy Efficiency and Sustainability}}
As LLMs are computationally intensive, energy consumption becomes a significant concern, especially in large-scale deployments. Future research should focus on improving the \textit{energy efficiency} of LLM-based frameworks. Techniques such as model compression, pruning, and the development of energy-efficient algorithms can reduce computational load and associated energy consumption.

  \renewenvironment{IEEEbiography}[1]
  {\IEEEbiographynophoto{#1}}
  {\endIEEEbiographynophoto}
  
\vspace*{-5\baselineskip}
\begin{IEEEbiography}{Asmaa Abdallah} received a Ph.D. in electrical engineering from the American University of Beirut, Beirut, Lebanon, in 2020. She is currently a research scientist at KAUST. 
\end{IEEEbiography}

\vspace*{-5\baselineskip}
\begin{IEEEbiography}{Abdullatif Albaseer} received a Ph.D. in Computer Science and Engineering from Hamad Bin Khalifa University (HBKU), Doha, Qatar, where he is currently a postdoctoral researcher.
\end{IEEEbiography}

\vspace*{-5\baselineskip}
\begin{IEEEbiography}{Abdulkadir Celik} received a Ph.D. in co-majors of electrical engineering and computer engineering from Iowa State University, Ames, IA, USA, in 2016. He is currently a senior research scientist at KAUST. 
\end{IEEEbiography}

\vspace*{-5\baselineskip}
\begin{IEEEbiography}{Mohamed Abdallah} received a Ph.D. degree in Electrical and Computer Engineering from the University of Maryland at College Park, College Park, MD, USA, in 2006.  He is currently a full professor at HBKU.
\end{IEEEbiography}

\vspace*{-5\baselineskip}
\begin{IEEEbiography}{Ahmed M. Eltawil} received a Ph.D. degree in electrical engineering from the University of California, Los Angeles, CA, USA, in 2003. He is currently a full professor at KAUST.
\end{IEEEbiography}


\begin{thebibliography}{10}
\bibitem{rong2024leveraging}
B.~Rong and H.~Rutagemwa, ``Leveraging large language models for intelligent
  control of {6G} integrated {TN}-{NTN} with iot service,'' \emph{IEEE
  Network}, 2024.

\bibitem{celik2024genai}
A.~Celik and A.~M. Eltawil, ``At the dawn of generative ai era: A
  tutorial-cum-survey on new frontiers in 6g wireless intelligence,''
  \emph{IEEE Open Journal of the Comms. Soc.}, vol.~5, pp. 2433--2489, 2024.

\bibitem{zou2023wireless}
H.~Zou \emph{et~al.}, ``Wireless multi-agent generative ai: From connected
  intelligence to collective intelligence,'' \emph{arXiv preprint
  arXiv:2307.02757}, 2023.

\bibitem{xu2024large}
S.~Xu \emph{et~al.}, ``Large multi-modal models ({LMM}s) as universal
  foundation models for {AI}-native wireless systems,'' \emph{arXiv preprint
  arXiv:2402.01748}, 2024.

\bibitem{shen2024large}
Y.~Shen, J.~Shao, X.~Zhang, Z.~Lin, H.~Pan, D.~Li, J.~Zhang, and K.~B. Letaief,
  ``Large language models empowered autonomous edge {AI} for connected
  intelligence,'' \emph{IEEE Commun. Mag.}, 2024, early access.

\bibitem{bariah2024large}
L.~Bariah \emph{et~al.}, ``Large generative ai models for telecom: The next big
  thing?'' \emph{IEEE Commun. Mag.}, 2024.

\bibitem{jiang2024large}
F.~Jiang, Y.~Peng, L.~Dong, K.~Wang, K.~Yang, C.~Pan, D.~Niyato, and O.~A.
  Dobre, ``Large language model enhanced multi-agent systems for 6g
  communications,'' \emph{IEEE Wireless Communications}, 2024.

\bibitem{shao2024wirelessllm}
J.~Shao \emph{et~al.}, ``Wireless{LLM}: Empowering large language models
  towards wireless intelligence,'' \emph{arXiv preprint arXiv:2405.17053},
  2024.

\bibitem{tarkoma2023ai}
S.~Tarkoma, R.~Morabito, and J.~Sauvola, ``{AI}-native interconnect framework
  for integration of large language model technologies in 6{G} systems,''
  \emph{arXiv preprint arXiv:2311.05842}, 2023.

\bibitem{6Gllm}
M.~Xu \emph{et~al.}, ``When large language model agents meet {6G} networks:
  Perception, grounding, and alignment,'' 2024.

\bibitem{yilma2024telecomrag}
G.~M. Yilma \emph{et~al.}, ``Telecom{RAG}: Taming telecom standards with
  retrieval augmented generation and llms,'' \emph{arXiv preprint
  arXiv:2406.07053}, 2024.

\bibitem{maatouk2023teleqna}
A.~Maatouk \emph{et~al.}, ``Tele{Q}n{A}: A benchmark dataset to assess large
  language models telecommunications knowledge,'' \emph{arXiv preprint
  arXiv:2310.15051}, 2023.

\bibitem{zou2024telecomgpt}
H.~Zou \emph{et~al.}, ``Telecom{GPT}: A framework to build telecom-specfic
  large language models,'' \emph{arXiv preprint arXiv:2407.09424}, 2024.

\bibitem{nazar2024enwar}
A.~M. Nazar, A.~Celik, M.~Y. Selim, A.~Abdallah, D.~Qiao, and A.~M. Eltawil,
  ``{ENWAR}: A {RAG}-empowered multi-modal {LLM} framework for wireless
  environment perception,'' \emph{arXiv preprint arXiv:2410.18104}, 2024.

\bibitem{PA2018debbah}
L.~Sanguinetti \emph{et~al.}, ``Deep learning power allocation in massive
  mimo,'' in \emph{2018 52nd Asilomar Conf. on Signals, Systems, and
  Computers}, 2018, pp. 1257--1261.

\end{thebibliography}
\end{document}